\documentclass[trackchanges]{aastex701}

\begin{document}


\title[]{Black Hole Mass Estimation of a Background Quasar in the Vicinity of the cD Galaxy Holm~15A}

\author[orcid=0000-0002-4436-221X]{Emmanuel R\'{i}os-L\'{o}pez}
\altaffiliation{Email: riloemm@inaoep.mx}
\affiliation{Instituto Nacional de Astrofísica, Óptica y Electrónica, Coordinación de Astrofísica, Tonantzintla, Puebla, México.}
\affiliation{Secretaría de Ciencias, Humanidades, Tecnología e Innovación, Av. Insurgentes Sur 1582, 03940, Ciudad de México, México.}
\email[]{riloemm@inaoep.mx}  

\author[orcid=0000-0002-1381-7437]{Omar López-Cruz}
\affiliation{Instituto Nacional de Astrofísica, Óptica y Electrónica, Coordinación de Astrofísica, Tonantzintla, Puebla, México.}
\email{omarlx@inaoep.mx}

\author[orcid=0000-0002-2733-0670]{Ana Luisa González-Morán}
\affiliation{Instituto Nacional de Astrofísica, Óptica y Electrónica, Coordinación de Astrofísica, Tonantzintla, Puebla, México.}
\affiliation{Secretaría de Ciencias, Humanidades, Tecnología e Innovación, Av. Insurgentes Sur 1582, 03940, Ciudad de México, México.}
\email[]{analuisagm@inaoep.mx}

\author[orcid=0000-0002-3721-8869]{Christopher Añorve}
\affiliation{Universidad Autónoma de Sinaloa, FACINE, Ciudad Universitaria s/n, Culiac\'an, Sinaloa, M\'exico.}
\email{canorve@uas.edu.mx}


\begin{abstract}
This study analyzes spectra from the central region of the galaxy cluster Abell 85 observed with GTC/OSIRIS, focusing on the cD galaxy Holm 15A and the point-like X-ray source SDSS J004150.75-091824.3, projected $13.74\arcsec$ from Holm 15A. For the latter, we found it is a background quasar with a redshift of $1.56386 \pm 0.00397$, ruling out the possibility of it being a rogue BH gravitationally bound to Holm 15A. In addition, using single-epoch spectroscopy, we report that the quasar J004150 hosts a Supermassive Black Hole (SMBH) with a mass of 
$\log(M_{\text{BH}}/\mathrm{M_{\odot}}) = 8.72 \pm 0.26$.
\end{abstract}



\section{INTRODUCTION} \label{Intro}
We present observations of the central region of the galaxy cluster Abell 85: the cD galaxy Holm~15A and SDSS J004150.75-091824.3 (hereafter, J004150). Holm~15A was reported by \citet{Lopez-Cruz_et_al_2014} as the galaxy with the largest core ($r_\gamma = 4.57 \pm 0.06\; \mathrm{kpc}$) and that could host an Ultramassive Black Hole (UMBH) with  $M_{BH} \sim 10^{10}\;\mathrm{M_{\odot}}$ (even $M_{BH} \sim$ $10^{11}\;\mathrm{M_{\odot}}$ with some scaling relations). But \cite{Mehrgan_et_al_2019} reported 
$\mathrm{M_{\text{BH}}}=4.0 \pm 0.01\times 10^{10}\;\mathrm{M_{\sun}}$ using the stellar velocity field.

J004150, with $z_{phot}\sim 0.9$ \citep{Richards_et_al_2009}, lies within the halo of Holm~15A. Then, \citeauthor{Lopez-Cruz_et_al_2014} proposed that J004150 may be a rogue SMBH ejected from the nucleus of Holm~15A ($z=0.05536 \pm 0.00002$) due to the scouring action of SMBH triples \citep[e.g.,][]{2007MNRAS.377..957H, 2022ApJ...940L..49N}. Also, the present study estimates the SMBH mass of J004150 using single-epoch virial mass estimators based on broad emission lines of active galactic nucleus.

\section{Data and Methodology} \label{Data-Met}

\begin{figure}
\begin{center}
\includegraphics[width=8.5 cm]{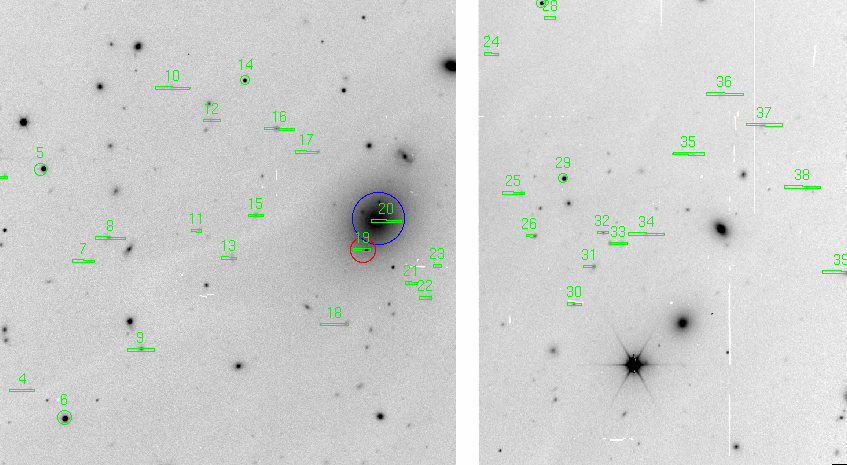}
\includegraphics[width=9.0 cm]{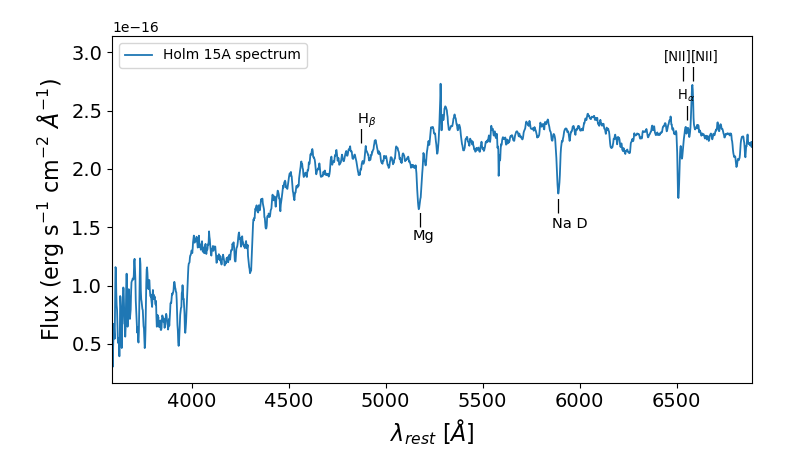}\\
\includegraphics[width=9.2 cm]{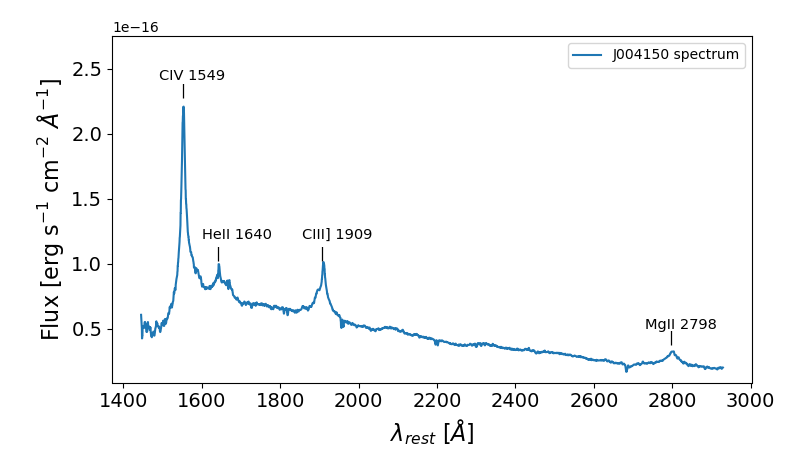}
\includegraphics[width=8.7 cm]{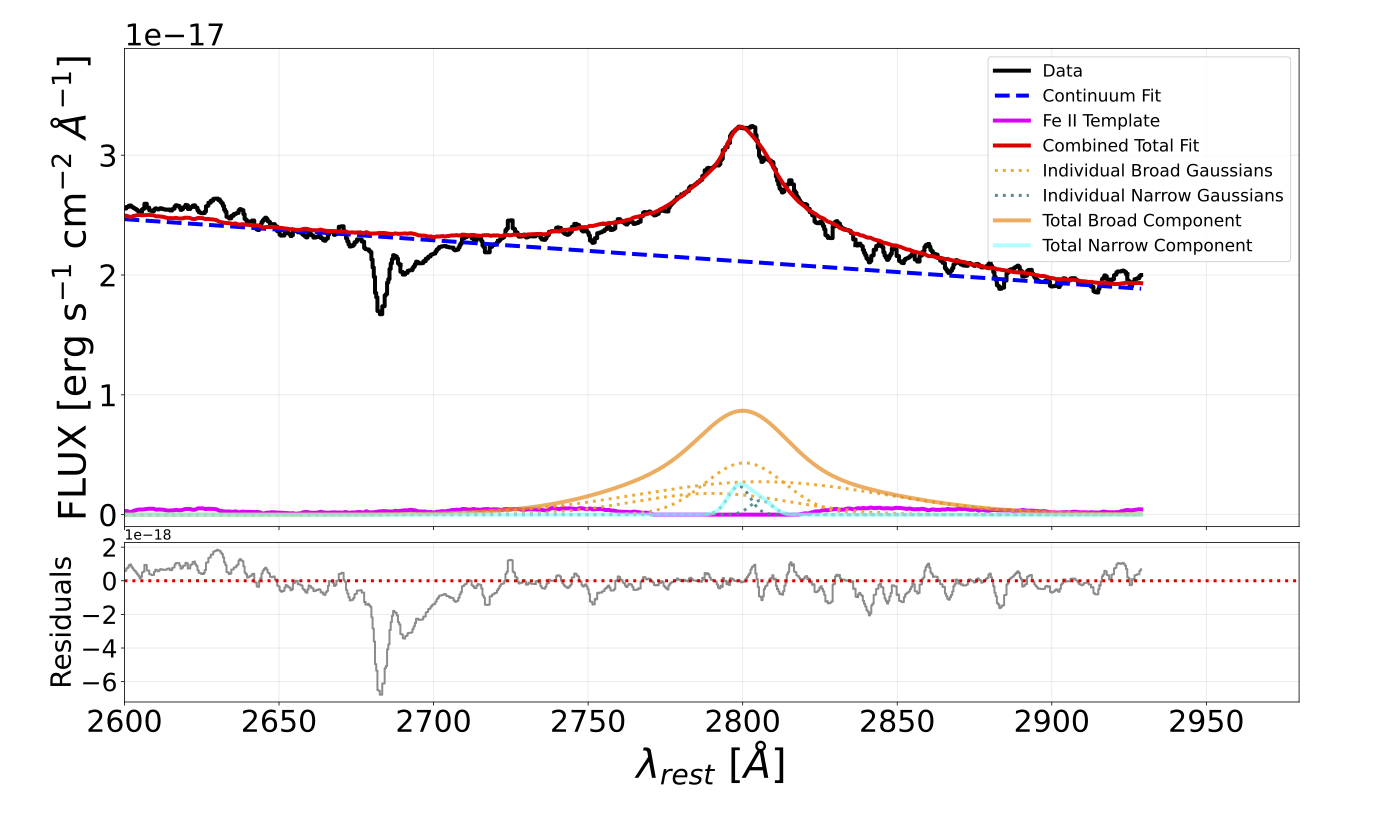}\\
\caption{\footnotesize Top-left: FoV of GTC/OSIRIS showing Holm 15A (blue circle), J004150 (red circle) and slit positions in green rectangles. 1D spectra of Holm 15A (top-right) and J004150 (bottom-left). Bottom-right: Spectral decomposition for the Mg II line of J004150 (top subpanel): original data in black, power-law continuum in blue and combined best-fit model (continuum + UV Fe II template + narrow-line component + broad-line component) in red. Individual broad-line Gaussians are the orange dashed lines and the total broad component is the orange solid line; individual narrow-line Gaussians are the green dashed lines and the total narrow component is the cyan line; Fe II template is the magenta line; lower subpanel shows residuals (gray) between the combined best-fit model and original data.}
\label{Fig1}
\end{center}
\end{figure}

We observed the central region of Abell~85 (top-left panel of Fig. \ref{Fig1}) using the  Gran Telescopio Canarias (GTC) equipped with the OSIRIS Multi-Object Spectrograph (MOS) under the programme GTC5-17AMEX (PI: ERL). We carried out the observations in service mode using the R1000B grism, slit width of $1.2\arcsec$, covering a spectral range from $\sim$3700-7500 \AA, with a spectral resolution of $\sim$11 \AA.

To estimate the BH mass of J004150, we used the BH virial mass estimator based on the Mg II $\lambda$2798  emission line:

\begin{equation}
\mathrm{log}\left(\frac{M_\mathrm{BH}}{M_{\odot}}\right)= a + b \, \mathrm{log}\left[\frac{L}{10^{44} \, \mathrm{erg \, s^{-1}}}\right] + c \, \mathrm{log}\left[\frac{\mathrm{FWHM}(\mathrm{Mg \, II})}{\mathrm{km \, s^{-1}}}\right],
\label{eq1}  
\end{equation}

\noindent where $L$ represents the continuum luminosity at $\lambda=3000$ \AA, $\lambda \, L_{\lambda}$, or Mg II line luminosity, L(Mg II); FWHM is the full width at half maximum of the MgII line; coefficients values are $a=1.933$, $b = 0.63$ and $c= 2.0$ for L(Mg II) \citep{Shen2012}, and $a=0.86$, $b= 0.5$ and $c= 2.0$ for $\lambda \, L_{\lambda}$  \citep{2009ApJ...699..800V}. To find these parameters, we performed the spectral decomposition of the Mg II line (bottom-right panel of Fig. \ref{Fig1}) similarly to \citet[][]{ 2017ApJ...843...30Z}. We fit a pseudo-continuum model consisting of a power-law and the UV Fe II template \citep{2005ApJ...625..688V}. We subtracted the pseudo-continuum from the spectrum and fitted the Mg II line with broad and narrow components in the [2690,2910] \AA\ range. We modeled the broad component with multiple Gaussians, using three Gaussians (each with FWHM $\ge$ 900 km/s); for the narrow component, we used two Gaussians and checked both possibilities of doublet ratio, (2:1) for optically thin and (1:1) for optically thick. We used Astropy \citep{2022ApJ...935..167A} to fit the continuum and the Mg II line profile. 

To estimate the uncertainties in the parameters of Eq. 1, we performed a Monte Carlo analysis in which we randomly perturbed the original spectrum to generate 50 mock spectra. The resulting values are $\lambda$ L$_{\lambda}$ = 0.9097 $\pm$ 0.0072 $\times 10^{45}$ erg/s, L(Mg II)= 8.720 $\pm$ 0.069 $\times 10^{42}$ erg/s and FWHM (Mg II) = 4896.9 $\pm$ 452.9 km/s. To estimate the uncertainty of the BH mass, we propagated the errors using Eq. \ref{eq1} as:

\begin{equation}    
\sigma_{\mathrm{log} M_{\mathrm{BH}}}=\sqrt{(\sigma_{int})^2 +(c\,\sigma_\mathrm{log FWHM})^2 + (b\, \sigma_{\mathrm{log}L})^2}
\label{eq2}  
\end{equation}

\noindent where $\sigma_{int}=0.25$ or 0.33 dex is the intrinsic scatter of the calibration from \cite{2009ApJ...699..800V} or \citet{Shen2012}, respectively; $\sigma_{\mathrm{log}L}$ is the uncertainty in $\lambda L_{\lambda}$ or L(Mg II) and $\sigma_\mathrm{log FWHM}$ is the FWHM(Mg II) uncertainty.\\

\section{Results and Discussion}
\label{Res}

We found that J004140 is a background quasar at $z=1.56386 \pm 0.00397$\footnote{This redshift was reported in the \href{https://smf.mx/programas/congreso-nacional-de-fisica/memorias-cnf/}{\textit{CNF 2018}} and \href{https://www.inaoep.mx/~progharo/gh2018/scientific_program.html}{\textit{GH Workshop 2018}} in the presentation \textit{Is There a Third Supermassive Black Hole Associated with the cD Galaxy Holm 15A?}}, measured using CIV $\lambda$1549, HeII $\lambda$1640, CIII] $\lambda$1909 and MgII $\lambda$2798 lines (bottom-left panel of Fig. \ref{Fig1}); our redshift results improved over \citet{2021PASP..133a4101M}. For Holm~15A, we report $ z = 0.05565 \pm 0.00052$ using $H_{\alpha}$, $H_{\beta}$, Mg, and Na D lines (top-right panel of Fig. \ref{Fig1}), in agreement with \citet{Lopez-Cruz_et_al_2014}. These results clearly rule out \citeauthor{Lopez-Cruz_et_al_2014}
suggestion that J004140 is a BH ejected from the nucleus of Holm~15A.

We obtained a SMBH mass for J004150 of 
$\log(M_{\text{BH}}/\mathrm{M_{\odot}}) = 8.72 \pm 0.26$ utilizing the \citet{2009ApJ...699..800V} calibration. 
Since our spectrum ends at $\lambda_{\text{rest}} \sim 2930\text{ \AA}$, which prevented a direct measurement of the continuum at $3000\text{\AA}$, we implemented an extrapolation according to \citet{Shen2011}; thus, we independently calculated the mass using the \citet{Shen2012} relation, obtaining \(\log(M_{\text{BH}}/\mathrm{M_{\odot}}) = 8.65 \pm 0.34\). The agreement between both estimates confirms the consistency in the SMBH masses.

Given recent observations reporting overmassive BHs at high redshift \citep[e.g.,][]{Chon_et_al_2026}, BH mass estimates in quasars at intermediate redshift become relevant for constraining evolution models of galaxies.


\begin{acknowledgments}
This work is based on observations using GTC at the Observatorio del Roque de los Muchachos of the IAC.\\

\end{acknowledgments}

\newpage
\bibliography{sample701}{}

@ARTICLE{2017ApJ...843...30Z,
       author = {{Zhu}, Dongchun and {Sun}, Mouyuan and {Wang}, Tinggui},
        title = "{Relation between the Variations in the Mg II {\ensuremath{\lambda}}2798 Emission Line and 3000 {\r{A}} Continuum}",
      journal = {\apj},
         year = 2017,
        month = jul,
       volume = {843},
       number = {1},
          eid = {30},
        pages = {30},
          doi = {10.3847/1538-4357/aa76e7},
archivePrefix = {arXiv},
       eprint = {1706.01635},
 primaryClass = {astro-ph.GA},
       adsurl = {https://ui.adsabs.harvard.edu/abs/2017ApJ...843...30Z}
}

@ARTICLE{2009ApJ...699..800V,
       author = {{Vestergaard}, M. and {Osmer}, Patrick S.},
        title = "{Mass Functions of the Active Black Holes in Distant Quasars from the Large Bright Quasar Survey, the Bright Quasar Survey, and the Color-selected Sample of the SDSS Fall Equatorial Stripe}",
      journal = {\apj},
         year = 2009,
        month = jul,
       volume = {699},
       number = {1},
        pages = {800-816},
          doi = {10.1088/0004-637X/699/1/800},
archivePrefix = {arXiv},
       eprint = {0904.3348},
 primaryClass = {astro-ph.CO},
       adsurl = {https://ui.adsabs.harvard.edu/abs/2009ApJ...699..800V}
}

@ARTICLE{2005ApJ...625..688V,
       author = {{Vestergaard}, Marianne and {Peterson}, Bradley M.},
        title = "{Variability of Fe II Emission Features in the Seyfert 1 Galaxy NGC 5548}",
      journal = {\apj},
         year = 2005,
        month = jun,
       volume = {625},
       number = {2},
        pages = {688-698},
          doi = {10.1086/429791},
archivePrefix = {arXiv},
       eprint = {astro-ph/0502433},
 primaryClass = {astro-ph},
       adsurl = {https://ui.adsabs.harvard.edu/abs/2005ApJ...625..688V}
}

@ARTICLE{2022ApJ...940L..49N,
       author = {{Ni}, Yueying and {Di Matteo}, Tiziana and {Chen}, Nianyi and {Croft}, Rupert and {Bird}, Simeon},
        title = "{Ultramassive Black Holes Formed by Triple Quasar Mergers at z   2}",
      journal = {\apjl},
         year = 2022,
        month = dec,
       volume = {940},
       number = {2},
          eid = {L49},
        pages = {L49},
          doi = {10.3847/2041-8213/aca160},
archivePrefix = {arXiv},
       eprint = {2209.01249},
 primaryClass = {astro-ph.GA},
       adsurl = {https://ui.adsabs.harvard.edu/abs/2022ApJ...940L..49N}
}

@ARTICLE{2007MNRAS.377..957H,
       author = {{Hoffman}, Loren and {Loeb}, Abraham},
        title = "{Dynamics of triple black hole systems in hierarchically merging massive galaxies}",
      journal = {\mnras},
         year = 2007,
        month = may,
       volume = {377},
       number = {3},
        pages = {957-976},
          doi = {10.1111/j.1365-2966.2007.11694.x},
archivePrefix = {arXiv},
       eprint = {astro-ph/0612517},
 primaryClass = {astro-ph},
       adsurl = {https://ui.adsabs.harvard.edu/abs/2007MNRAS.377..957H}
}

@ARTICLE{2022ApJ...935..167A,
       author = {{Price-Whelan}, Adrian M. and {Lim}, Pey Lian and {Earl}, Nicholas and {Starkman}, Nathaniel and {Bradley}, Larry and {Shupe}, David L. and {Patil}, Aarya A. and {Corrales}, Lia and {Brasseur}, C.~E. and {N{\"o}the}, Maximilian and {Donath}, Axel and {Tollerud}, Erik and {Morris}, Brett M. and {Ginsburg}, Adam and {Vaher}, Eero and {Weaver}, Benjamin A. and {Tocknell}, James and {Jamieson}, William and {van Kerkwijk}, Marten H. and {Robitaille}, Thomas P. and {Merry}, Bruce and {Bachetti}, Matteo and {G{\"u}nther}, H. Moritz and {Aldcroft}, Thomas L. and {Alvarado-Montes}, Jaime A. and {Archibald}, Anne M. and {B{\'o}di}, Attila and {Bapat}, Shreyas and {Barentsen}, Geert and {Baz{\'a}n}, Juanjo and {Biswas}, Manish and {Boquien}, M{\'e}d{\'e}ric and {Burke}, D.~J. and {Cara}, Daria and {Cara}, Mihai and {Conroy}, Kyle E. and {Conseil}, Simon and {Craig}, Matthew W. and {Cross}, Robert M. and {Cruz}, Kelle L. and {D'Eugenio}, Francesco and {Dencheva}, Nadia and {Devillepoix}, Hadrien A.~R. and {Dietrich}, J{\"o}rg P. and {Eigenbrot}, Arthur Davis and {Erben}, Thomas and {Ferreira}, Leonardo and {Foreman-Mackey}, Daniel and {Fox}, Ryan and {Freij}, Nabil and {Garg}, Suyog and {Geda}, Robel and {Glattly}, Lauren and {Gondhalekar}, Yash and {Gordon}, Karl D. and {Grant}, David and {Greenfield}, Perry and {Groener}, Austen M. and {Guest}, Steve and {Gurovich}, Sebastian and {Handberg}, Rasmus and {Hart}, Akeem and {Hatfield-Dodds}, Zac and {Homeier}, Derek and {Hosseinzadeh}, Griffin and {Jenness}, Tim and {Jones}, Craig K. and {Joseph}, Prajwel and {Kalmbach}, J. Bryce and {Karamehmetoglu}, Emir and {Ka{\l}uszy{\'n}ski}, Miko{\l}aj and {Kelley}, Michael S.~P. and {Kern}, Nicholas and {Kerzendorf}, Wolfgang E. and {Koch}, Eric W. and {Kulumani}, Shankar and {Lee}, Antony and {Ly}, Chun and {Ma}, Zhiyuan and {MacBride}, Conor and {Maljaars}, Jakob M. and {Muna}, Demitri and {Murphy}, N.~A. and {Norman}, Henrik and {O'Steen}, Richard and {Oman}, Kyle A. and {Pacifici}, Camilla and {Pascual}, Sergio and {Pascual-Granado}, J. and {Patil}, Rohit R. and {Perren}, Gabriel I. and {Pickering}, Timothy E. and {Rastogi}, Tanuj and {Roulston}, Benjamin R. and {Ryan}, Daniel F. and {Rykoff}, Eli S. and {Sabater}, Jose and {Sakurikar}, Parikshit and {Salgado}, Jes{\'u}s and {Sanghi}, Aniket and {Saunders}, Nicholas and {Savchenko}, Volodymyr and {Schwardt}, Ludwig and {Seifert-Eckert}, Michael and {Shih}, Albert Y. and {Jain}, Anany Shrey and {Shukla}, Gyanendra and {Sick}, Jonathan and {Simpson}, Chris and {Singanamalla}, Sudheesh and {Singer}, Leo P. and {Singhal}, Jaladh and {Sinha}, Manodeep and {Sip{\H{o}}cz}, Brigitta M. and {Spitler}, Lee R. and {Stansby}, David and {Streicher}, Ole and {{\v{S}}umak}, Jani and {Swinbank}, John D. and {Taranu}, Dan S. and {Tewary}, Nikita and {Tremblay}, Grant R. and {de Val-Borro}, Miguel and {Van Kooten}, Samuel J. and {Vasovi{\'c}}, Zlatan and {Verma}, Shresth and {de Miranda Cardoso}, Jos{\'e} Vin{\'\i}cius and {Williams}, Peter K.~G. and {Wilson}, Tom J. and {Winkel}, Benjamin and {Wood-Vasey}, W.~M. and {Xue}, Rui and {Yoachim}, Peter and {Zhang}, Chen and {Zonca}, Andrea and {Astropy Project Contributors}},
        title = "{The Astropy Project: Sustaining and Growing a Community-oriented Open-source Project and the Latest Major Release (v5.0) of the Core Package}",
      journal = {\apj},
         year = 2022,
        month = aug,
       volume = {935},
       number = {2},
          eid = {167},
        pages = {167},
          doi = {10.3847/1538-4357/ac7c74},
archivePrefix = {arXiv},
       eprint = {2206.14220},
 primaryClass = {astro-ph.IM},
       adsurl = {https://ui.adsabs.harvard.edu/abs/2022ApJ...935..167A}
}

@ARTICLE{Lopez-Cruz_et_al_2014,
   author = {{L{\'o}pez-Cruz}, O. and {A{\~n}orve}, C. and {Birkinshaw}, M. and 
	{Worrall}, D.~M. and {Ibarra-Medel}, H.~J. and {Barkhouse}, W.~A. and 
	{Torres-Papaqui}, J.~P. and {Motta}, V.},
    title = "{The Brightest Cluster Galaxy in A85: The Largest Core Known So Far}",
  journal = {The Astrophysical Journal Letters},
archivePrefix = "arXiv",
   eprint = {1405.7758},
     year = 2014,
    month = nov,
   volume = 795,
      eid = {L31},
    pages = {L31},
      doi = {10.1088/2041-8205/795/2/L31},
   adsurl = {http://adsabs.harvard.edu/abs/2014ApJ...795L..31L}
}

@ARTICLE{Mehrgan_et_al_2019,
       author = {{Mehrgan}, Kianusch and {Thomas}, Jens and {Saglia}, Roberto and {Mazzalay}, Ximena and {Erwin}, Peter and {Bender}, Ralf and {Kluge}, Matthias and {Fabricius}, Maximilian},
        title = "{A 40 Billion Solar-mass Black Hole in the Extreme Core of Holm 15A, the Central Galaxy of Abell 85}",
      journal = {\apj},
         year = 2019,
        month = dec,
       volume = {887},
       number = {2},
          eid = {195},
        pages = {195},
          doi = {10.3847/1538-4357/ab5856},
archivePrefix = {arXiv},
       eprint = {1907.10608},
 primaryClass = {astro-ph.GA},
       adsurl = {https://ui.adsabs.harvard.edu/abs/2019ApJ...887..195M}
}

@ARTICLE{Richards_et_al_2009,
       author = {{Richards}, Gordon T. and {Myers}, Adam D. and {Gray}, Alexander G. and {Riegel}, Ryan N. and {Nichol}, Robert C. and {Brunner}, Robert J. and {Szalay}, Alexander S. and {Schneider}, Donald P. and {Anderson}, Scott F.},
        title = "{Efficient Photometric Selection of Quasars from the Sloan Digital Sky Survey. II. \raisebox{-0.5ex}\textasciitilde1,000,000 Quasars from Data Release 6}",
      journal = {\apjs},
         year = 2009,
        month = jan,
       volume = {180},
       number = {1},
        pages = {67-83},
          doi = {10.1088/0067-0049/180/1/67},
archivePrefix = {arXiv},
       eprint = {0809.3952},
 primaryClass = {astro-ph},
       adsurl = {https://ui.adsabs.harvard.edu/abs/2009ApJS..180...67R}
}

@ARTICLE{Chon_et_al_2026,
       author = {{Chon}, Sunmyon and {Hirano}, Shingo and {Ishiyama}, Tomoaki and {Chang}, Seok-Jun and {Springel}, Volker},
        title = "{Overmassive black holes and little red dots naturally form in simulations}",
      journal = {\nat},
         year = 2026,
        month = sep,
       volume = {657},
       number = {8132},
        pages = {621-625},
          doi = {10.1038/s41586-026-10985-8},
       adsurl = {https://ui.adsabs.harvard.edu/abs/2026Natur.657..621C}
}
\bibliographystyle{aasjournalv7}



\end{document}